\documentclass[12pt,preprint]{aastex}

\bibpunct[, ]{(}{)}{;}{a}{}{,}
\newcommand{\swift}{{\it Swift}}
\newcommand{\xrt}{{\rm XRT}}
\newcommand{\grb}{GRB~$050509$B}
\newcommand{\cluster}{ZwCl~1234.0+02916}

\slugcomment{Revised ApJL version, October 4, 2005}

\shorttitle{\grb\ host cluster}
\shortauthors{Pedersen et al.}

\begin{document}

\title{The host galaxy cluster of the short gamma-ray burst GRB~050509B%
\thanks{Based on observations collected at the European
Southern Observatory, Paranal, Chile (ESO Programme 075.D-0261).}
}

\author{K.~Pedersen\altaffilmark{2},
\'{A}.~El\'{\i}asd\'ottir\altaffilmark{2}, 
J.~Hjorth\altaffilmark{2}, 
R.~Starling\altaffilmark{3},
J.~M.~Castro~Cer\'on\altaffilmark{2},
J.~P.~U.~Fynbo\altaffilmark{2}, 
J.~Gorosabel\altaffilmark{4},
P.~Jakobsson\altaffilmark{2},
J.~Sollerman\altaffilmark{2,5},
D.~Watson\altaffilmark{2}}

\altaffiltext{2}{Dark Cosmology Centre, Niels Bohr Institute, University of
Copenhagen, Juliane Maries Vej 30, DK-2100 Copenhagen \O, Denmark; 
kp, ardis, jfynbo, pallja, darach @astro.ku.dk, josemari@alumni.nd.edu.}
\altaffiltext{3}{Astronomical Institute `Anton Pannekoek', University of 
Amsterdam, 1098 SJ Amsterdam, The Netherlands; starling@science.uva.nl}
\altaffiltext{4}{Instituto de Astrof\'{\i}sica de Andaluc\'{\i}a (IAA-CSOC,
P.O. Box 03004, E-18080 Granada, Spain; jgu@iaa.es}
\altaffiltext{5}{Department of Astronomy, Stockholm University, AlbaNova,
106 91 Stockholm, Sweden; jesper@astro.ku.dk}

\begin{abstract}
The first arcsecond localization of a short gamma-ray burst, \grb, 
has enabled detailed studies of a short burst environment. 
We here report on studies of the environment of \grb\ using
the {\em Swift} X-ray Telescope (XRT). The \xrt\ error circle of the burst 
overlaps with an
elliptical galaxy in the cluster of galaxies \cluster. Based on the measured
X-ray flux of the cluster we estimate that the probability for a chance 
superposition of \grb\ and a cluster at least as X-ray bright
as this cluster is $< 2\times 10^{-3}$, presenting the first strong case of a
short burst located in a cluster of galaxies.
We also consider the case for \grb\ being located behind \cluster\ 
and gravitationally lensed. From the velocity dispersion of
the elliptical galaxy and the temperature of hot intracluster gas, 
we model the mass distribution in the elliptical galaxy and the 
cluster, and calculate the gravitational lensing magnification within
the \xrt\ error circle. 
We find that, if GRB050509B would be positioned significantly behind the
cluster, it is most likely magnified by a factor less than two, 
but that the burst could be strongly lensed if it is positioned within 
$2\arcsec$ of the center of the bright elliptical galaxy.
Further mapping of arcsecond size short burst error boxes is a new promising 
route to determine the spatial distribution of old stars throughout the
Universe.
\end{abstract}


\section{INTRODUCTION}

About 25\% of all gamma-ray bursts detected by BATSE were ``short'' 
\citep[duration less than 2~s,][]{chryssa} and had hard spectra. 
Due to the non-detection of an afterglow from a short burst, 
rendering quick and accurate 
localization unfeasible, their nature have remained elusive.
\grb\ was detected by the \swift\ Burst Alert Telescope (BAT)
on 2005~May~9 at 04:00:19.23 (UT) \citep{gehrels,bloom} and upon
slewing the \swift\ X-Ray Telescope (XRT) started observations 
of the burst 62~s after the BAT was triggered. For the first time, 
an X-ray afterglow from a short burst was detected enabling
a localization within $9.3\arcsec$ \citep{gehrels}.
Imaging of the \xrt\ error box showed that it overlapped with an
elliptical galaxy with a redshift of $z=0.2248$ \citep{bloom}.
Based on the unlikeliness of a chance alignment between \grb\ and such
a galaxy it has 
been argued that this is the host galaxy of the burst \citep{bloom,gehrels}.
However, several much fainter (and presumably more distant) 
galaxies were also detected within the \xrt\ error circle 
\citep[see Fig.1 in][]{hjorth05}.
A unique feature of \grb\ is that it is situated in the direction of a 
nearby cluster of galaxies, identified in the Zwicky catalog 
\citep[\cluster,][]{zwicky}
and in the Northern Sky Optical Cluster Survey with a photometric 
redshift of $z=0.2214$ \citep[NSC~J123610+285901,][]{gal}.
X-ray emission from a hot intracluster medium in \cluster\ is clearly 
detected by
the \xrt\ from which \citet{bloom} determined the cluster X-ray centroid 
and temperature.
If \grb\ is hosted by one of the fainter galaxies beyond the cluster
the emission from the burst is inevitably boosted by gravitational lensing. 
This could enhance the probability of detecting the burst
and imply a (possibly large) correction to the derived burst energetics.

Here we (i) derive the X-ray flux, temperature and mass of the cluster 
\cluster\
and estimate the probability of a chance alignment of a gamma-ray burst and
a cluster with an X-ray flux at least as large as that of \cluster, and 
(ii) calculate
the gravitational lensing magnification within the \grb\ \xrt\ error circle, 
constraining scenarios where \grb\ is gravitationally lensed. 
A cosmology with $H_0=70$~km~s$^{-1}$~Mpc$^{-1}$, $\Omega_m=0.3$, 
$\Omega_\Lambda=0.7$ is assumed throughout this Letter. All errors quoted
are 90\% confidence limits.

\section{OBSERVATIONS AND DATA REDUCTION}
The \swift\ \xrt\ \citep{xrt} observed the \grb\ region for a duration of 75~ks. 
The X-ray afterglow of \grb\ is detected during 
the first 400~s of the \xrt\ observation. We investigated the 
X-ray emission from an intracluster medium in the cluster \cluster, 
by analyzing \xrt\ data obtained after the first 400~s where the \grb\ 
X-ray afterglow had faded below detection. Only data taken in 
``photon counting mode'',
providing full imaging and spectroscopic information, were included
in the analysis yielding 31.9~ks of exposure time. 
Using the {\em xrtpipeline} procedure version 0.8.8 in the HEAsoft 6.0 software
package\footnote{http://heasarc.gsfc.nasa.gov/docs/software/lheasoft} 
we applied standard screening parameters and the latest available 
calibration files (2005 June 1) to the Level~1 data 
obtained from the \swift\ Quick-Look Data 
compilation\footnote{http://swift.gsfc.nasa.gov/cgi-bin/sdc/ql?, 
data sequence 00118749000} to produce cleaned event lists of the full \xrt\
field of view. Subsequently, the {\em xselect} tool in the HEAsoft 6.0 package
was used to produce images and spectra with the spatial and spectral filters
given below.

An image was obtained in the 0.3--10~keV band and sources were detected 
using a wavelet algorithm ({\em wavdetect} implemented in the Chandra Interactive
Analysis of Observations (CIAO)\footnote{http://cxc.harvard.edu/ciao} version
3.2 software package). The astrometry was fixed in the following way: We
measured the centroid of the X-ray afterglow in the \swift\ frame of reference 
from the first 400~s of the \xrt\ observations. Then the X-ray afterglow centroid 
was adjusted to match the position determined by \citet{gehrels} and \citet{burrowsgcn} 
(who registered 
a 50~ks {\em Chandra} image to 2MASS coordinates, and then registered a 30~ks \xrt\ 
image to the {\em Chandra} image). Hence we apply a shift to the \swift\ reference 
frame of $\Delta$R.A.(J2000)$=-2{\farcs}57$ and $\Delta$decl.(J2000)$=-2{\farcs}16$. 

In order to characterize the intracluster emission,
two point sources near the cluster were excluded and the source regions were
filled with a Poissonian pixel value distribution sampled from the surroundings
(using the {\em dmfilth} procedure in CIAO 3.2). 
Finally, we ran the {\em wavdetect} procedure on the resulting image for
determining the centroid of the intracluster emission.

X-ray spectra were generated from events in the 0.5--7~keV range
within a circular region centered on the cluster X-ray centroid and 
with a radius between $1.75\arcmin$ and $2.17\arcmin$ (corresponding to
the radius at which the intracluster emission is $1.5$ 
times above the background level or equals the background level, respectively). 
The energy bounds
were chosen so as to leave out the lowest energies where the spectral
calibration is most uncertain, and to disregard the highest energies 
dominated by background. Background spectra were generated from circular 
regions devoid of sources, not overlapping with the cluster extraction
region or each other, and with the same radius as used for the 
intracluster emission. The average background level is $\sim 10\%$ of 
the intracluster emission. We used the {\em Xspec} package version 12.2.0 
\citep{xspec}
to fit background subtracted spectra of the intracluster emission,
binned to a minimum of 20~counts per bin, with a hot plasma 
model \citep[MEKAL,][]{mewe,lied} absorbed by neutral matter along the line of sight.
The Galactic absorbing column density in this direction 
is $N_{\rm H}\sim 1.52\times 10^{20}$~cm$^{-2}$ \citep{dicklock}.

\section{CLUSTER X-RAY PROPERTIES}

\begin{figure}
\plotone{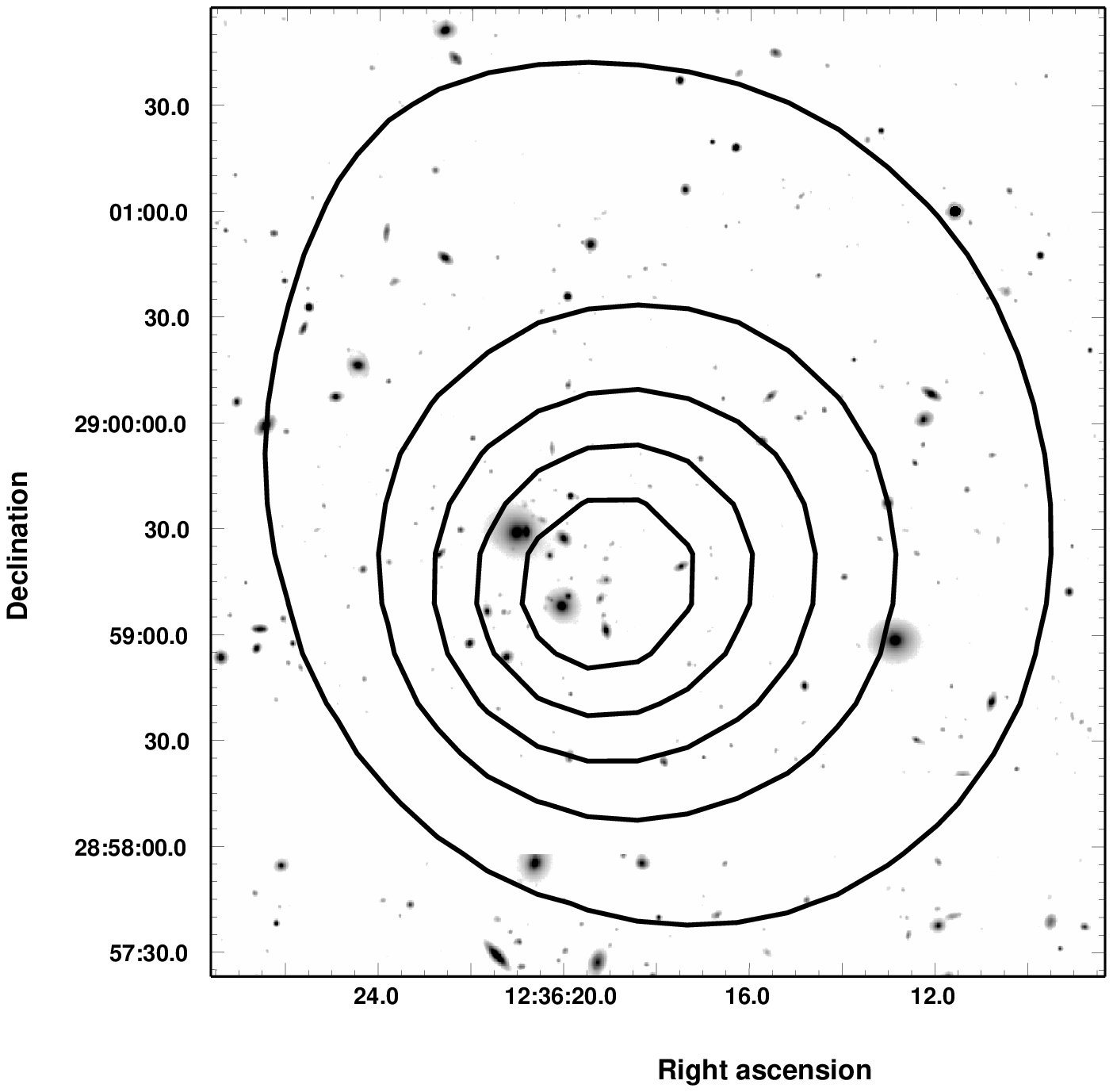}
\caption{X-ray contours from adaptively smoothed \xrt{} 0.3--10~keV 
image with point sources removed overlaid on 
VLT~FORS2~$R$-band image of \cluster\ \citep{hjorth05}. The outermost contour 
($7.0\times 10^{-6}$~counts~arcsec$^{-2}$~s$^{-1}$) corresponds to 1.5
times the background,
and the contours increase consecutively by a factor 1.5. The region between
declinations $+28{\arcdeg}57{\arcmin}58{\arcsec}$ and 
$+28{\arcdeg}58{\arcmin}20{\arcsec}$ 
is not covered due the FORS2 chip gap.\label{fig1}}
\end{figure}

The intracluster X-ray emission centroid is 
R.A.(J2000)$=12^{\rm h} 36^{\rm m} 19.03^{\rm s}$,\\ 
decl.(J2000)$=+28\arcdeg 59\arcmin 13\arcsec$ (with an uncertainty of $\sim 5\arcsec$).
This is $72\arcsec$ off-set from the center of the \xrt\ error circle.
The diffuse emission extends to about $2\arcmin$ corresponding to
430~kpc for a cluster redshift of $z=0.2248$ \citep{bloom}.
327~net counts (0.3--10~keV) are detected within $2.17\arcmin$ of the X-ray
centroid. With the spatial resolution achieved by the photon statistics the 
morphology of the intracluster emission is close to circular with a slight
NE-SW elongation in the cluster outskirts (see Fig.\ref{fig1}).
Observations obtained with the {\em Chandra X-ray Observatory} can further
reveal the structure of the intracluster medium \citep{patel}.
The near coincidence between the X-ray centroid and several bright 
elliptical galaxies strongly suggests 
that this is the bottom of the cluster potential well rather than the 
optical center of ZwCl~1234.0+2916 
\citep[R.A.(J2000)$=12^{\rm h} 36^{\rm m} 28.0^{\rm s}$, 
decl.(J2000)$=+28\arcdeg 59\arcmin 30\arcsec$,][]{zwicky} or NSC~J123610+285901
\citep[R.A.(J2000)=$12^{\rm h} 36^{\rm m} 10.2^{\rm s}$, 
decl.(J2000)=$+28\arcdeg 59\arcmin 01\arcsec$,][]{gal}. 
At the cataloged optical cluster centers no enhancement in X-ray 
emission is seen.

\begin{figure}
\includegraphics[angle=270,scale=0.6]{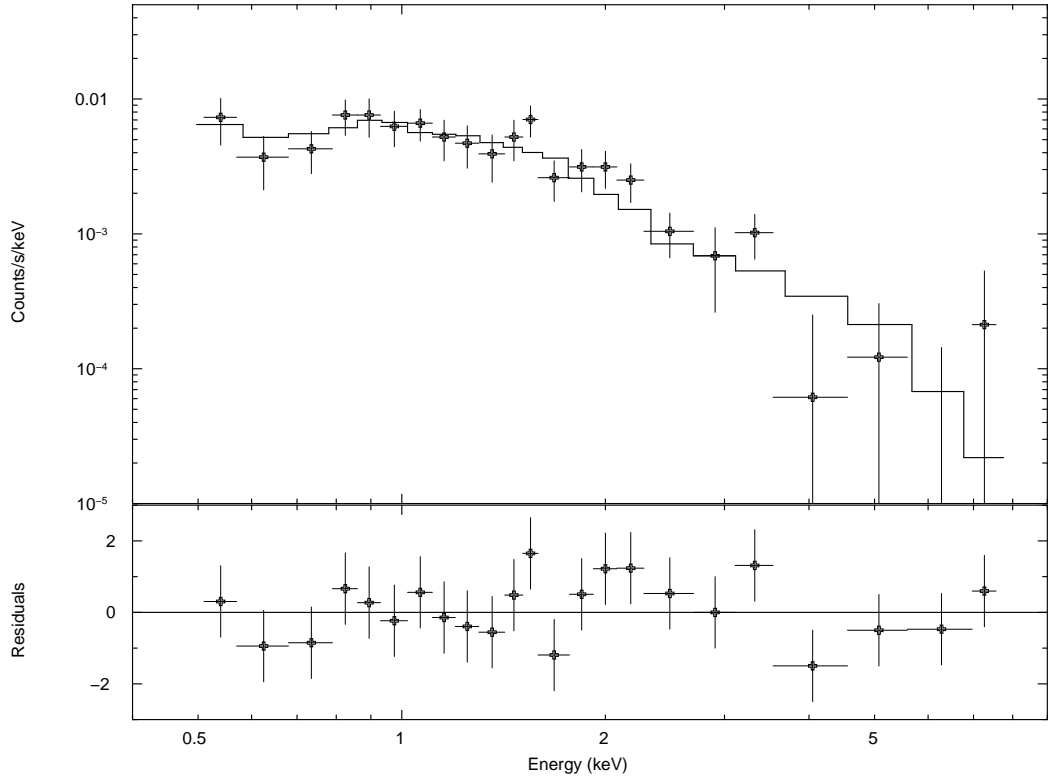}
\caption{Top panel: \xrt\ X-ray spectrum of the cluster emission with best fit absorbed
MEKAL hot plasma model. Bottom panel: Residuals from best fit in units of standard
deviations.\label{fig2}}
\end{figure}

The spectrum of the intracluster medium within a radius of $2.17\arcmin$
of the X-ray centroid is well fitted ($\chi^2=15.8$ for 21 d.o.f.) by a MEKAL 
model with absorption fixed at the Galactic value and a typical abundance of 
$Z=0.25$~$Z_{\sun}$ 
\citep{ml}, see Fig.~\ref{fig2}.
The best fit temperature is $kT=3.65^{+2.26}_{-1.22}$~keV and the 0.5--2~keV flux
$1.21^{+0.33}_{-0.37}\times 10^{-13}$~erg~cm$^{-2}$~s$^{-1}$.
The main systematic error in determining the temperature and flux is the background 
subtraction. Using four different background regions results in best fit temperatures
between $kT=3.01^{+1.84}_{-0.83}$~keV and $kT=3.78^{+2.71}_{-1.26}$~keV, and
0.5--2~keV fluxes between 
$0.96^{+0.52}_{-0.66}\times 10^{-13}$~erg~cm$^{-2}$~s$^{-1}$ and 
$1.34^{+0.50}_{-0.57}\times 10^{-13}$~erg~cm$^{-2}$~s$^{-1}$.

We also investigated the influence on the best temperature from varying (i) the
cluster emission extraction radius, (ii) the absorbing column density, 
and (iii) the intracluster abundance. 
Fitting the spectrum extracted within a radius of $1.75\arcmin$ of the cluster X-ray
centroid yields a temperature of $kT=3.79^{+2.77}_{-1.21}$~keV.
Letting the absorbing column density be a free parameter in the fit results in a 
best fit value of $N_{\rm H}=7.5^{+12.27}_{-7.5}\times 10^{20}$~cm$^{-2}$ and 
the best fit temperature is $kT=2.90^{+2.30}_{-1.00}$~keV. However, given that
only a small amount of extra-galactic absorption is expected \citep{allen}, 
and that the Galactic absorbing column
density is 50\% uncertain in worst case, a maximum column density of 
$N_{\rm H}=2.28\times 10^{20}$~cm$^{-2}$ is expected. For this column density,
the best fit temperature is $kT=3.54^{+2.32}_{-1.16}$~keV.
The abundance is constrained to be $Z<1.5$~$Z_{\sun}$ and varying the 
abundance in the range $0.2-0.3$~$Z_{\sun}$ 
\citep[corresponding to the observed abundance variation in clusters,][]{ml}
changes the best fit temperature by only $\sim 1\%$. 

From the above analysis we conclude that the global temperature of the \cluster\
intracluster medium is $kT=3.65^{+2.26+0.15}_{-1.22-0.64}$~keV
where the latter uncertainty is due to systematics from background 
subtraction. Within the errors this is consistent with the temperature derived by 
\citet{bloom} ($kT=5.25^{+3.36}_{-1.68}$~keV).
For a cluster redshift of $z=0.2248$ the 0.1--2~keV luminosity is 
$(2.5\pm 0.7)\times 10^{43}$~erg~s$^{-1}$ which is in good agreement with
the expected luminosity from the empirical luminosity--temperature relation 
of clusters \citep{popesso}.
The X-ray properties of \cluster\ thus shows it to be a cluster intermediate 
to the Virgo and Coma clusters.

The cluster mass can be estimated from an empirical mass--X-ray temperature relation.
Using the X-ray determined mass--temperature relation of \citet{arnaud} for their
full sample we find $M_{500}=2.2^{+3.3}_{-0.6}\times 10^{14}$~M$_{\sun}$ 
taking into account the uncertainties in the mass--temperature relation and the 
uncertainties in the temperature. $M_{500}$ is the mass within the radius at which 
the mean cluster density is 500~times the critical density of the Universe. 
From the optically determined mass--temperature relation of \citet{popesso}
we find $M_{500}=2.3^{+3.1}_{-1.2}\times 10^{14}$~M$_{\sun}$ in excellent agreement
with the above mass estimate.

\section{GRAVITATIONAL LENSING MAGNIFICATION}
The alignment of the \grb\ \xrt\ error circle and a bright elliptical 
galaxy, G1, in the cluster \cluster\ makes it relevant to consider
the possibility that \grb\ is a gravitationally lensed background source
\citep{es}.
Hence, we have calculated the gravitational lensing magnification from G1 and the 
cluster for several possible redshifts of \grb.

The gravitational lensing calculations were carried out using the 
{\em gravlens} software package \citep{keeton}.
We model the galaxy as a singular isothermal ellipsoid with an axis 
ratio of 0.81 \citep{bloom} with the semimajor axis aligned along a 
position angle $90\degr$, a velocity dispersion of 
$260\pm 40$~km~s$^{-1}$ \citep{bloom}, and the cluster as a singular 
isothermal sphere with $M_{500}=2.2_{-0.6}^{+3.3}\times 10^{14}$~M$_{\sun}$
(see Section 3). We take the center of the \xrt\ error circle to be at 
R.A.(J2000)$=12^{\rm h}36^{\rm m}13{\fs}58$ and 
decl.(J2000)$=+28{\degr}59{\arcmin}01{\farcs}3$ 
\citep{gehrels} and choose that as the center of our coordinate system. 
We place the center of the galaxy G1 at R.A.(J2000)=$12^{\rm h}36^{\rm m}12{\fs}86$ 
and decl.(J2000)$=+28{\degr}58{\arcmin}58{\farcs}0$ and the center of the cluster at 
the X-ray centroid of the diffuse cluster emission (see Section 3).

Below we give results for point sources at the expected lower and upper
limit on the \grb\ redshift $z=1.3$ and $z=3$, respectively. 
Due to the blue continuum and 
lack of emission lines in the faint galaxies in the \xrt\ error box, 
the lower limit on their redshift is $z \approx 1.3$ \citep{bloom}. 
The upper limit on the redshift is motivated by models predicting an
intrinsic duration of a GRB longer than 8~ms \citep{lee}.
We note that chosing a larger upper limit does not affect our conclusions.

\begin{figure}
\plotone{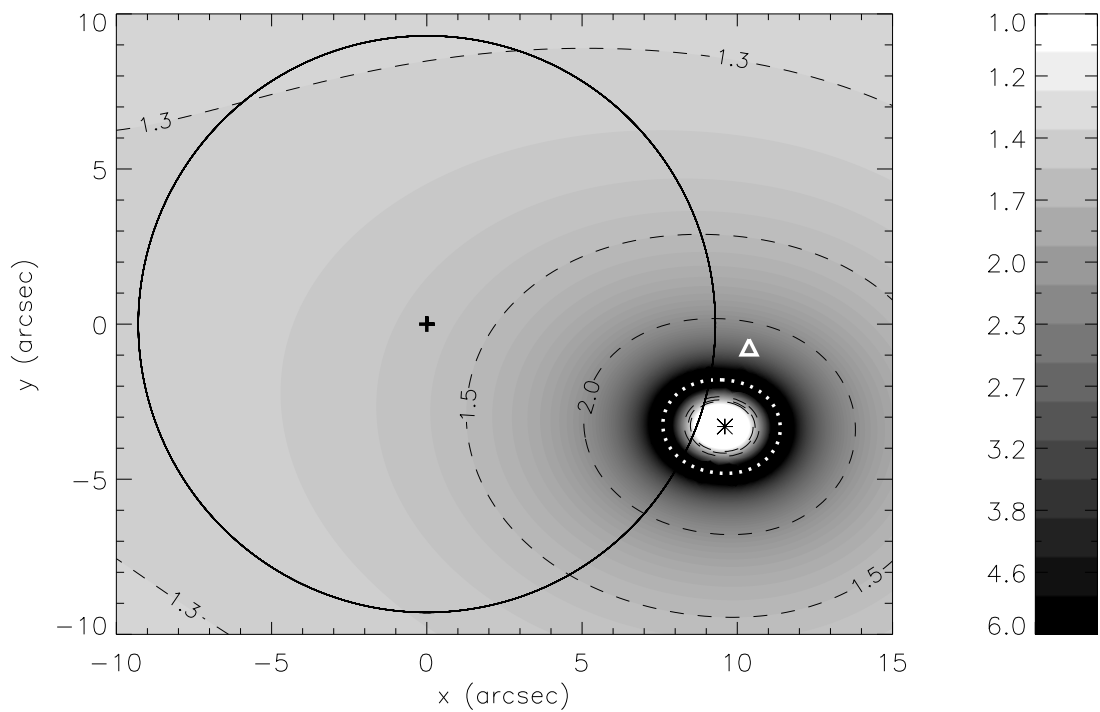}
\caption{Gravitational lensing magnification in and around the \xrt\ error circle
(centered on the black cross) shown as logarithmic gray scale (intensity scale
on the right) and contours (dashed lines). North is up and East to the left.
The source was placed at a redshift
$z=3$, and the mass distribution of the galaxy G1 and the cluster \cluster\
were modelled as isothermal ellipsoids with masses determined from the velocity
dispersion and the X-ray temperature, respectively.
The black star is the center of the galaxy G1. The white triangle 
marks the location of the source revealed by subtracting a model of galaxy G1
\citep{hjorth05}, and the white, dotted line is a critical curve.
\label{fig3}}
\end{figure}

Taking uncertainties in the
velocity dispersion of G1 into account this gives an Einstein radius 
for G1 of $b=1.67^{+0.55}_{-0.47}$~arcsec for $z=3$ and
$b=1.50^{+0.49}_{-0.42}$~arcsec for $z=1.3$.
Similarly we get for the cluster (taking the uncertainty in the cluster mass
into account) that $b=9.7^{+8.2}_{-1.8}$~arcsec for 
$z=3$ and $b=9.0^{+7.6}_{-1.7}$~arcsec for $z=1.3$.
We then calculate the median magnification within the \xrt\ error circle of 
radius $9.3{\arcsec}$ \citep{gehrels}, see Fig.~\ref{fig3}. 
Most of the magnification is provided by 
the galaxy G1, and the main effect of the cluster is to increase the magnification
in the East-West direction.

The \xrt\ error circle crosses the critical curves (paths in the image 
plane where a point source will be exposed to infinite magnification) for 
nearly all mass models of G1 and the cluster.
The median of the magnification within the error circle is
a factor of $1.3^{+0.4}_{-0.1}$ for $z=3$ and 
$1.3^{+0.3}_{-0.1}$ for $z=1.3$.
We thus find that potential the magnification of \grb\ is typically a factor 1--2, 
fairly independent of the redshift of \grb. However, if \grb\ is positioned
within $\sim 2\arcsec$~of~G1 it could be more strongly magnified.
We note that the source revealed by
subtracting away G1 \citep{hjorth05}, situated $2.68\arcsec$ from the center
of G1, is not a strongly lensed background source.

\section{DISCUSSION}

From our gravitational lensing study we conclude that lensing is not strongly 
boosting the probability that \grb\ is situated in a background galaxy.
Furthermore, \citet{bloom} estimate that the probability of a chance 
alignment between \grb\ and G1 is $\sim 5\times 10^{-3}$ (using the apparent 
magnitude of G1) while \cite{gehrels} arrive at probability of
$\sim 1\times 10^{-4}$ (using the luminosity and distance of G1). Interestingly,
a correlation between error boxes of short bursts in the BATSE catalog
and the positions of galaxies in the local Universe has been claimed \citep{tanvir}.

We now proceed to estimate the probability of a chance alignment between 
the \xrt\ error circle and a cluster of galaxies with an X-ray flux 
and optical richness 
at least as large as that of \cluster. The sky surface density of clusters 
to a 0.5--2~keV flux limit of the \cluster\ flux is well determined to 
0.5--1~clusters~deg$^{-2}$ from several X-ray cluster surveys \citep{rosati}. 
The probability for a chance alignment within $75\arcsec$ of the cluster
center and the \grb\ error circle center is equal to the sky coverage fraction 
of such clusters with a radius of $75\arcsec$, i.e.\ $0.7-1.3\times 10^{-3}$. 
The sky surface density of clusters in the 
Northern Sky Optical Cluster Survey with a richness at least as large
as NSC~J123610+285901 ($N_{\rm gals}=24.3$, corresponding to less than
Abell richness class 0) is $\sim 1.2$~deg$^{-2}$ \citep{gal}, yielding a
chance alignment probability of $\sim 1.6\times 10^{-3}$.
Although a posteriori statistics is always uncertain \grb\ comprises the
first strong case for a short burst located in a galaxy cluster, based on either 
the X-ray flux or the optical richness of \cluster.
The association of 
\grb\ with \cluster\ provides independent evidence that \grb\ is also associated
with the galaxy G1 since this is a cluster member. 
Interestingly, it has recently been suggested that the line-of-sight towards 
the short burst GRB~050813 coincides with a cluster at redshift $z=0.72$ 
\citep{gladders,berger}. Previously, studies of the few small error boxes 
of short bursts have not revealed any short burst associated with a cluster 
\citep{hurley,nakar} 

A statistical correlation between gamma-ray burst error boxes (of short as well as long
bursts) and the position of Abell clusters has been claimed \citep{kolatt,struble}, 
but subsequent refined analyses have failed to confirm this \citep{gorosabel,hurley99}.
However, Abell clusters are relatively nearby and rich so a correlation between 
short bursts and Abell clusters is only expected if short bursts occur predominantly
within the completeness redshift of Abell clusters 
\citep[$z\lesssim 0.15$,][]{ebeling}. The likely association between \grb\ and
the cluster \cluster\ motivates further studies of the spatial correlation
between galaxy clusters and short burst sky positions. Since \cluster\ is less
rich than Abell clusters it would be beneficial to use cluster catalogs 
including poorer and more distant clusters. Such studies would provide interesting
constraints on the environments of short bursts and hence on their origin.

If, as suggested by the favoured models \citep{piran}, short bursts originate from 
merging of compact stellar remnants their progenitors are $10^6-10^9$~yr old 
\citep{voss}. Targeting the environment of short bursts is thus potentially 
a new independent way for tracing the old stellar population throughout
the Universe. Up to now, entire populations of old stars in galaxies
have been identified through their integrated optical/near-infrared colors.
Short bursts are promising pointers to sites of individual old stars,
revealing the spatial distribution of old stars, be they located in cluster 
ellipticals, field galaxies, or even in intergalactic space \citep{fuk}.

\acknowledgments

The Dark Cosmology Centre is supported by the DNRF.
KP, JP, and JMCC acknowledge support from Instrument Centre for Danish Astrophysics,
and JMCC acknowledges support from NBI's International Ph.D. School of Excellence.
JG acknowledges is partially supported by the Spanish Ministry of
Science and Education through programmes ESP2002-04124-C03-01 and AYA2004-01515
(including FEDER funds). This work was supported by the European Community's Sixth 
Framework
Marie Curie Research Training Network Programme, Contract No. MRTN-CT-2004-505183
``ANGLES''.
The authors acknowledge benefits from
collaboration within the EC FP5 Research Training Network ``Gamma-Ray
Bursts -- An Enigma and a Tool''.

Facilities: \facility{VLT(FORS1,FORS2)}, \facility{Swift(XRT)}

\clearpage

\clearpage

\end{document}